# Binder-free CNT cathodes for Li-O$_2$ batteries with more than one life

Zeliang Su[1,2], Israel Temprano[3*], Nicolas Folastre[1,2], Victor Vanpeene[4],
Julie Villanova[4], Gregory Gachot[1,2], Elena Shevchenko[5],
Clare P. Grey[3], Alejandro A. Franco[1,2,6,7], Arnaud Demortiere[1,2,7*]

[1]Laboratoire de Réactivité et Chimie des Solides (LRCS), CNRS UMR 7314, Université de Picardie Jules Verne, Hub de l'Energie, Rue Baudelocque, 80039 Amiens Cedex, France.

[2]Réseau sur le Stockage Electrochimique de l'Energie (RS2E), CNRS FR 3459, Hub de l'Energie, Rue Baudelocque, 80039 Amiens Cedex, France.

[3]Yusuf Hamied Department of Chemistry, University of Cambridge, CB2 1EW, Cambridge, United Kingdom

[4]ID16B ESRF-The European Synchrotron, CS 40220 Grenoble Cedex 9 38043, France

[5]Argonne National Laboratory, Center for Nanoscale Materials, Argonne, Illinois 60439, USA

[6]Institut Universitaire de France, 101 Boulevard Saint Michel, 75005 Paris, France.

[7]ALISTORE-European Research Institute, CNRS FR 3104, Hub de l'Energie, Rue Baudelocque, 80039 Amiens Cedex, France.

**Corresponding Authors:**

Arnaud Demortière ✉: arnaud.demortiere@cnrs.fr

and Israel Temprano ✉: it251@cam.ac.uk

**Abstract**

Li-O$_2$ batteries (LOB) performance degradation ultimately occurs through the accumulation of discharge products and irreversible clogging of the porous electrode during the cycling. Electrode binder degradation in the presence of reduced oxygen species can result in additional coating of the conductive surface, exacerbating capacity fading. Herein, we establish a facile method to fabricate free-standing, binder-free electrodes for LOBs in which multi-wall carbon nanotubes (MWCNT) form cross-linked networks exhibiting high porosity, conductivity, and flexibility. These electrodes demonstrate high reproducibility upon cycling in LOBs. After cell death, efficient and inexpensive methods to wash away the accumulated discharge products are demonstrated, as reconditioning method. The second life usage of these electrodes is validated, without noticeable loss of performance. These findings aim to assist in the development of greener high energy density batteries while reducing manufacturing and recycling costs.

## 1. Introduction

The global warming associated with increasing concentrations of heat-trapping greenhouse gases in the Earth's atmosphere increases the urgency to develop and apply greener and more sustainable energy applications[1]. The electrification of the transport and portable electronics sectors, with lithium battery (LIB) technology as a fundamental element, is expanding at high rate, spearheading a broader energy transition towards a fossil fuel-free society. Global LIB production has grown from 75 GWh in 2011 to 400 GWh in 2020 and is expected to exceed 1400 GWh in 2025, accounting for the giga-factories under construction in major economies alone[2,3]. However, based on LIB technology and electric vehicle (EV) market analysis, the Ni and Co demand forecast for 2030 is expected to reach 2.5 times their global production capacity of the year 2016. Recently concerns have been raised about the shortage of these elements due to their geolocation rarity[4,5].

Simultaneously, there are growing concerns over the disposal/recycling of batteries, as commercial LIB typically have an average lifespan of 8-10 years[6]. Millions of tons of LIBs across the world are expected to exit the market by 2040, increasing pressure for developing end-of-life treatment large-scale processes in a sustainable manner[7–9]. Energy density and coulombic efficiency are no longer the only considerations for developing new LIB formulations. Durability, lifespan, and efficient end-of-life treatment becomes increasingly important aspects of further developments in battery research[10–12].

Li-air batteries (LABs) are a potential high-density energy storage system for many novel applications, thanks to their outstanding theoretical energy density (>2920 Wh/kg)[13,14] and relatively low environmental impact compared to LIBs (being transition-metal free). The development of LABs (currently Li-$O_2$ batteries (LOB), strictly speaking, as most experimental work is performed with pure $O_2$) is however, currently hindered by cell irreversibility, mainly due to progressive accumulation of discharge products on the air electrode. As electrochemical round-trips progress with relatively low coulombic efficiency, discharge and side products [15–20] accumulate, and the available surface at the electrode diminish[21], ultimately leading to cell failure. Furthermore, typical polymeric electrode binders, such as Poly(vinylidene fluoride)(PVDF), suffer degradation in the presence of reduced oxygen species resulting in additional coating of the conductive surface, exacerbating capacity fading . Strategies to solve or mitigate these limitations proposed in the literature range from (a) grafting catalysts or doped cathode materials [22–26]; (b) using redox mediators[27–29]; (c) stabilizing the conductive lithium superoxide as discharge product[20,30]; to (d) promoting alternative electrochemistry (such as LiOH), which produces fewer parasitic reactions[31–33]. With these challenges at the forefront of LAB research, key aspects of end-of-life treatment and material recyclability are lagging in the literature.

This work presents a simple and scalable way of preparing self-standing and binder-free air electrodes for LABs, lending them amenable for recycling at end-or-life, using efficient and costless methods.  Thus, highly porous electrodes, based on coiled multiwalled carbon nanotubes (MWCNT) are fabricated, tested, and recycled. The second life capabilities of these electrodes are tested in fresh batteries. We show that the electrochemical performance of the second life material is comparable to the pristine electrodes. The free-standing and binder-free electrodes can be either directly reconditioned or re-dispersed to form new electrodes prior to assembly in new cells. These findings open new routes towards efficient recovery of LAB electrodes for continuous recycling, increasing the potential of LABs as environmentally friendly and low-cost alternative form of energy storage system.

## 2. Results

### 2.1. Electrode fabrication and microstructural characterization

The as-purchased MWCNT powder is shown in Fig 1a, with nanotube arrays having diameters of 5 to 30 nm and lengths averaging 100 µm. As-received MWCNT powder was initially dispersed in isopropanol and then vacuum filtered, forming a homogeneous disc (Fig 1b). The first filtrate is often of dark color due to a small amount of CNT percolation through the high porosity of the glassy fiber filter, but the filtrate rapidly turns clear, further percolation stopped by the first formed stack of CNTs on the filter paper. It is worth noting that the solvent (isopropanol) can be reused in this process, reducing the total amount of solvent required in the manufacture (and recycling) of these electrodes. In the resulting highly flexible and smearable free-standing disc of entangled CNTs the nanotubes are bundled, as shown by SEM and digital images in Fig 1b and c.

This facile preparation method can produce binder-free electrodes of large dimensions and can be straightforwardly scaled up both in surface area or in thickness, by simply using a bigger funnel (*e.g.*, 7 cm diameter disc shown in Fig 1b) or extending the filtration time (see Figure S1 SEMs in supporting material). Throughout the current study, the thickness of the electrode is controlled between 75 to 120 mm, and the electrodes are cut into 1.27 cm diameter discs for electrochemical testing (see Experimental Methods).

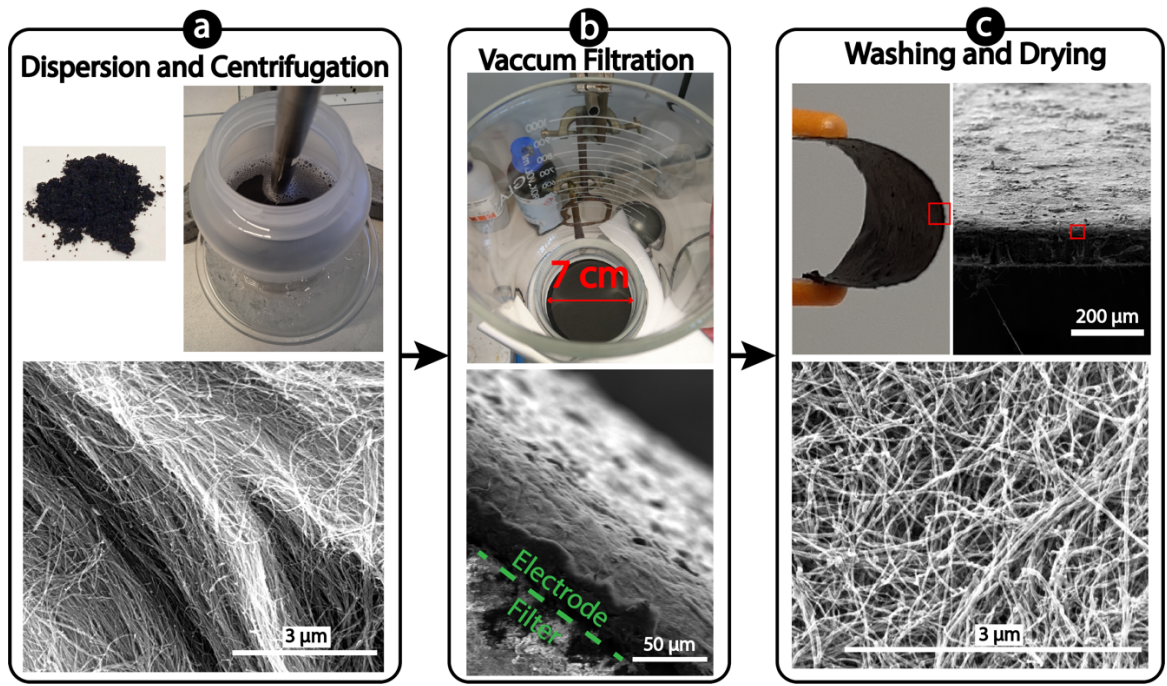

**Figure 1. Electrode preparation workflow.** a) Optical images of MWCNT powder and ultrasonic probe used for dispersion, and SEM image of the aligned array CNT powder; b) Optical image of a CNT disc on a vacuum filter and SEM image of the filtered material; c) Optical image showcasing the flexibility of the CNT discs and SEM images at different magnifications.

The 3D nano-architecture of the so-formed CNT electrodes were characterised using the synchrotron X-ray holo-tomography (nano-CT) technique at the ID16B beamline at the ESRF as described in our previous study [36]. Figure 2 shows the clear differentiation obtained between the MWCNT and the pores. From this data, the meso/macro porosity and tortuosity of this material were evaluated at 49% and 1.95, respectively, using the SegmentPy[37] and Taufactor[38,39] softwares (see Fig S2 for the segmentation justification and the representativeness calculation). These porosity and tortuosity values are consistent with those from $Li-O_2$ binder carbon electrode obtained by FIB/SEM 3D imaging [35]. The porosity of the electrode has a critical impact on the reaction intermediates at the surface while the tortuosity affects the electrochemical kinetics enhancing Li-ion diffusion inside the wetted porous electrode. The pore network distribution of over 400k pores detected in this volume was obtained using Porespy algorithm[40]. It is depicted in Fig 2c with a ball-and-stick model, showing pores and channels respectively. Fig 2d shows the statistical analysis of the porosity from the extracted pore network, indicating that the structure consists in dense small pores and short connections. The values extracted, *i.e.* average 180 nm in pore diameter and 80 nm in interconnections, revealed a highly connected carbon network with pore diameter close to that found for carbon super-P porous electrode in our previous work[22].

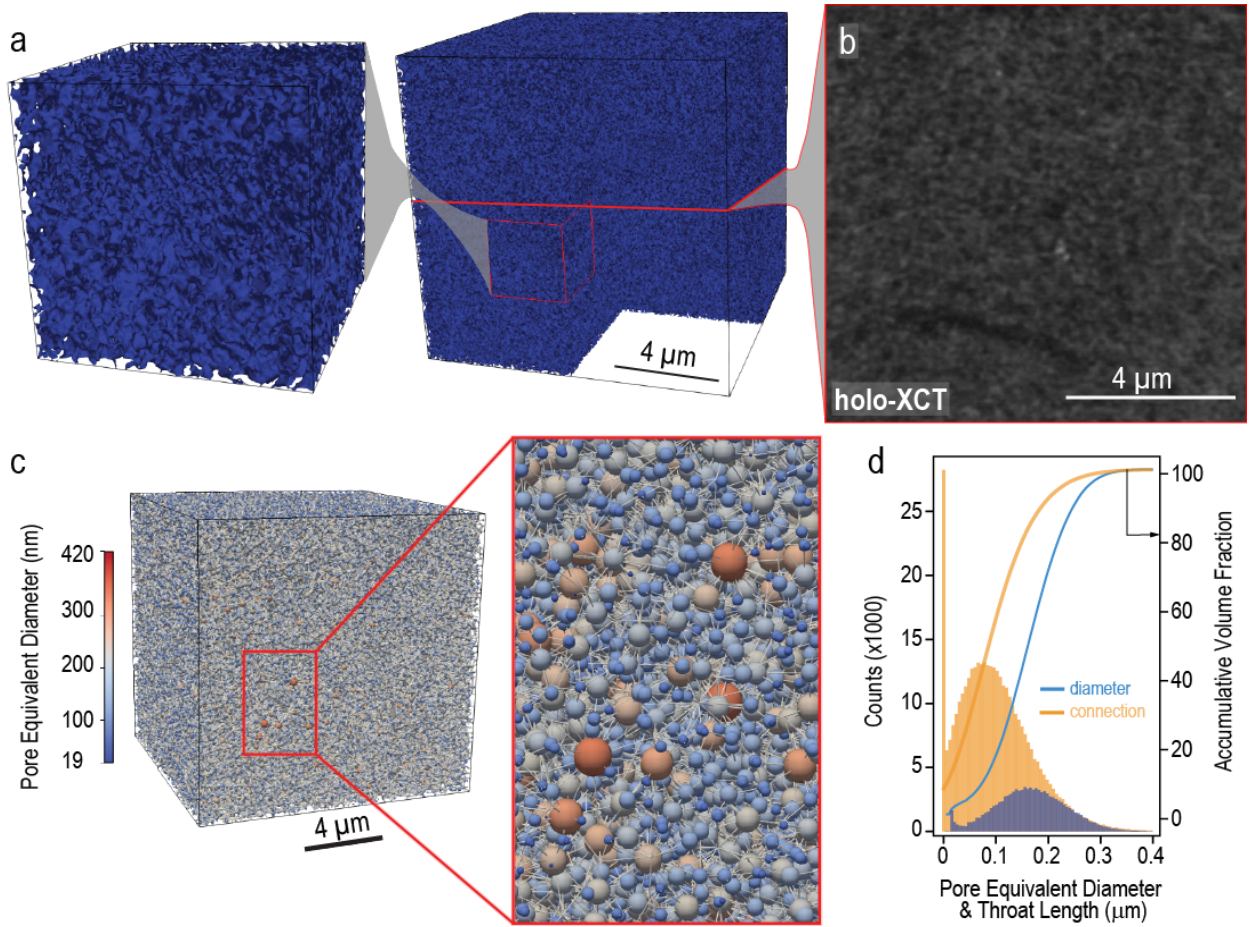

**Figure 2. Nano-CT data and structural analysis of binder-free CNT electrodes.** a) 3D render of the internal structure of pristine CNT electrodes; b) Cross-sectional slice of the nano-CT data showing the crisscrossed nanotubes; c) extracted pore-network model of the porous internal structure, where the nodes/balls represent the pores (radii are proportional to pore size, warmer color and larger size of balls indicate larger pores). The sticks represent the interconnections of these pores (many pores/balls are too small to render); d) the pore equivalent diameter and pore-interconnection distributions of the c) volume

## 2.2.   Electrochemical performance

A comparison of stability toward lithium metal for various electrolytes was performed to select the appropriate electrolyte (Fig 3a). Li|Li symmetric cells were cycled in commonly reported electrolytes in the LOB literature. Amongst these, 2M LiTFSI + 1M LiNO$_3$ in dimethylacetamide (DMA) showed the most stable behaviour, as previously reported by Yu *et al.*[43], and thus it was selected as baseline electrolyte for the remainder of this work. Li-O$_2$ cells with CNT air electrodes and this electrolyte, were tested to determine the total capacity (on deep discharge) and capacity retention evaluation (capacity limited to 500 mAh/g). Deep discharge tests showed capacities of ~1480 mAh/g$_{CNT}$ with a cut-off potential of 2 V (Fig 3b). The stoichiometry of the discharge and charge processes were evaluated using a Swagelok-based pressure monitoring system [41] (detailed in Figure S2). During the first discharge, a plateau can be observed at 2.7 V, corresponding to an oxygen reduction process (ORR) with stoichiometry of 2.62 e$^-$/O$_2$, as shown in figure 3a. Such deviation from the expected 2 e⁻/O$_2$ stoichiometry for Li$_2$O$_2$ formation would suggest either a substantial amounts of parasitic reactions, or a mixture of discharge products.

SEM post-mortem characterization of the air electrodes after discharge shows an abundance of large toroidal particles (SEM image in Fig 3c) indicative of Li$_2$O$_2$ crystals (red square), alongside regions of platelet structures (blue square), reminiscent of small LiOH crystals[32]. XRD data in Fig 3d suggests that the main crystalline product of the discharge is indeed lithium peroxide. The Rietveld analysis indicate that around 10% of the crystalline phases correspond to lithium hydroxide. FTIR/ATR spectra (Fig 3e), shows a feature assigned to the $\nu_{O-H}$ mode at 3676 cm$^{-1}$, further confirming the presence of LiOH in discharged electrodes. The presence of LiOH would explain the deviation from the 2e⁻/O$_2$ stoichiometry expected for Li$_2$O$_2$ formation during the ORR, as electrochemical LiOH formation is believed to occur through a 4e⁻/O$_2$ stoichiometry[32,33]. LiOH may form from trace amounts of water (<100 ppm of water measured by Carl Fischer titration) in the electrolyte (see Experimental Methods), or a catalytic effect of the electrode matrix[42]. Overall, the discharge capacity obtained during the discharge process of the CNT electrodes is in accordance with the literature[43].

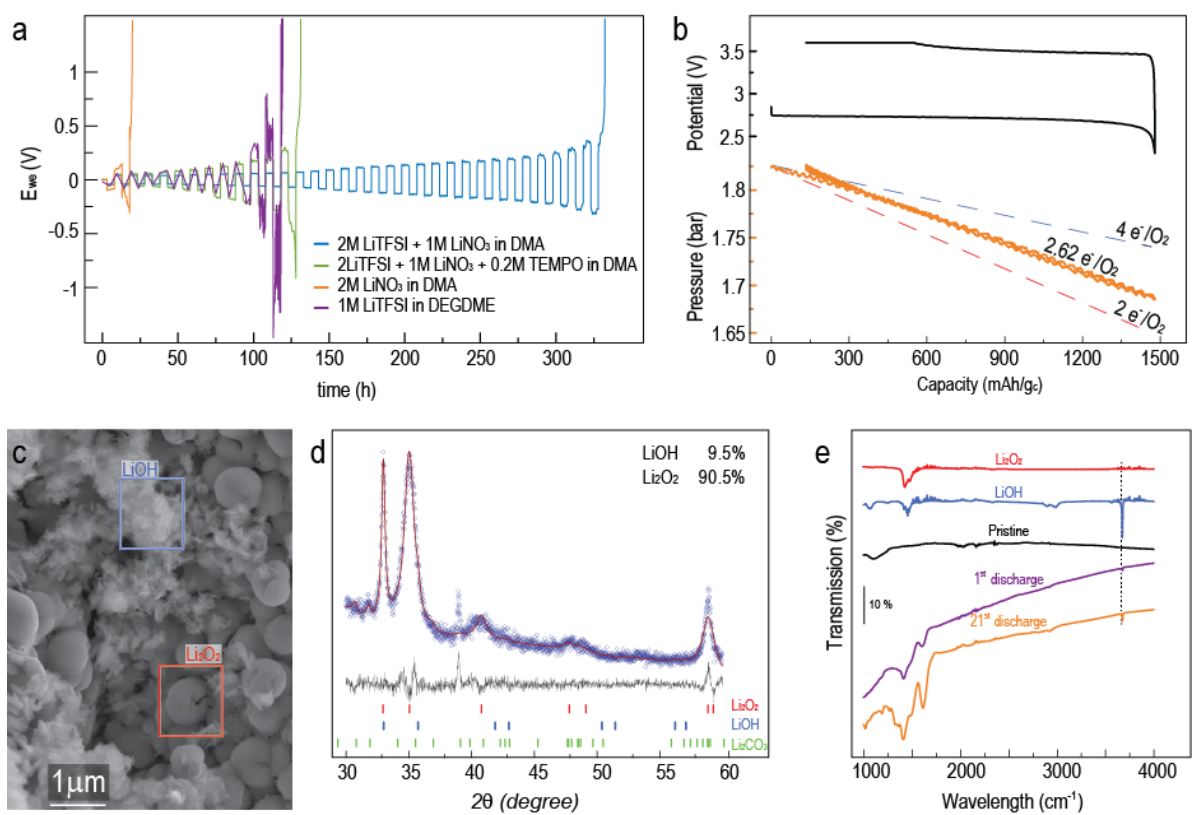

**Figure 3. Electrochemistry and discharge products analysis of the first deep discharge.** a) Li|Li symmetric cells (200µA/cm²) containing dimethylacetamide (DMA) with 2M of LiTFSI and 1M of LiNO$_3$ (blue); with 0.2M TEMPO added (green); 2M LiNO$_3$ (orange); and diethylene glycol dimethyl ether (DEGDME) with 1M LiTFSI (purple); b) Electrochemical (black) and pressure monitoring (orange) curves of the 1st deep discharge/charge cycle; c) SEM image; d) XRD and e) FTIR spectra from the air electrode after deep discharge.

During charge, the stoichiometry of the oxygen evolution reaction (OER) is also 2.62 e⁻/O$_2$ up to 3.6 V (Fig 3b), which suggests a highly reversible formation/decomposition of a mixture of discharge products with different stoichiometries (e.g., Li$_2$O$_2$ and LiOH)[32,33]. Several reports in the literature indicate that hydroxide (in addition to Li$_2$O$_2$) formation could be a reversible electrochemical process at relatively low overpotentials due to catalytic effects of either redox mediators [31,43], solid catalysts[44], or even hydrophilic CNT-based materials[42].

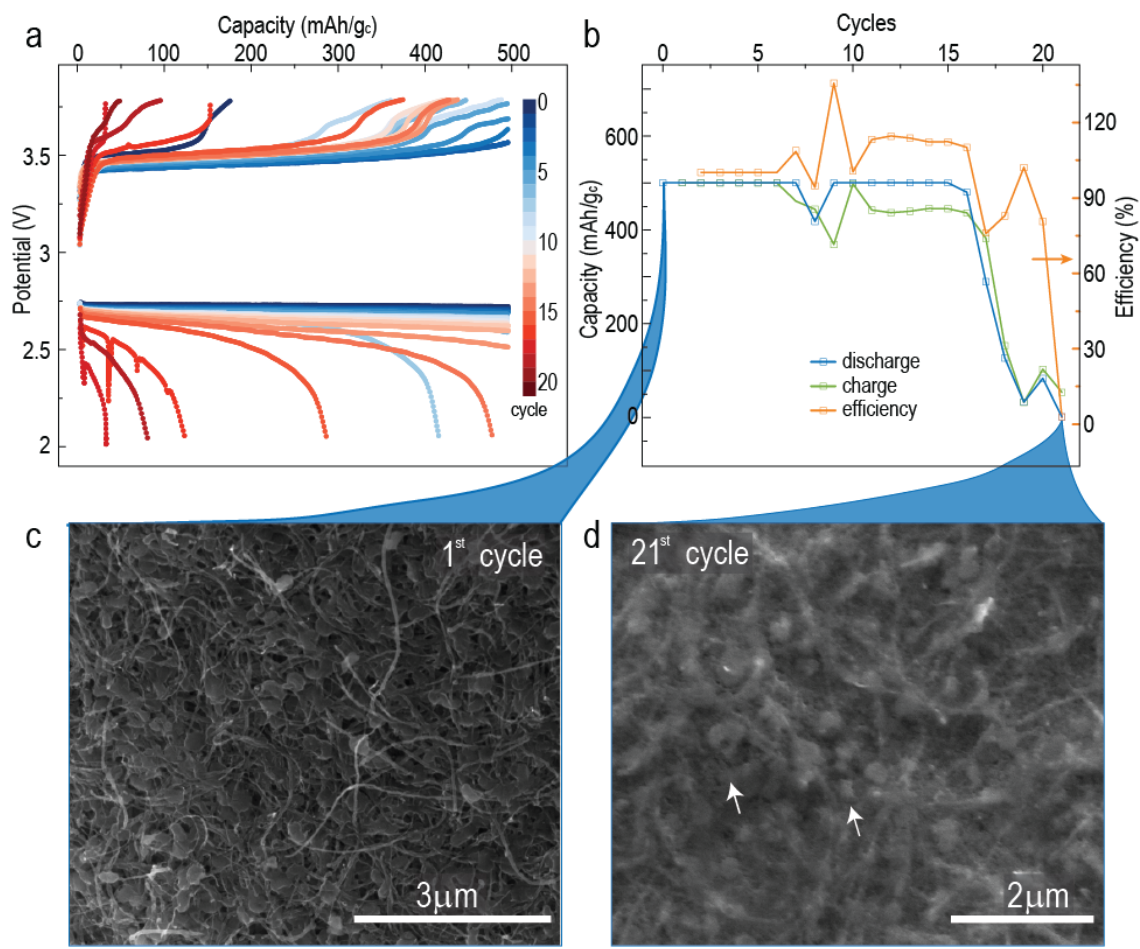

**Figure 4. Electrochemical data and SEM images.** a) Voltage-capacity profiles of 20 cycles; b) discharge/charge capacities and coulombic efficiency; c) SEM image of the electrode after the first discharge; d) SEM image of the electrode after 20 cycles.

In the case of deep discharge cycling (potential limited), a galvanostatic charge process can only recover 60% of the discharge capacity up to 3.6 V before a steep potential slope occurs (Fig S3). Allowing for the cell potential to rise above 3.6 V results in a reduction in slope of the pressure signal (Fig S3), indicating that the Oxygen Evolution process (OER) is limited beyond 3.6 V in these cells. To accommodate for the slow kinetics of the OER process at the end of charge, a CC-CV (constant current - constant voltage) protocol, with a potential hold at 3.6 V was thus used (Fig S4) increasing the faradaic efficiency, and thus capacity recovery whilst limiting the degradation that electrodes can suffer at high potentials.

The CNT-based cells were tested over multiple cycles to a limited capacity of 500 mAh/g between 2 and 4 V at 40 mA/g. The characteristic oxygen reduction plateaus at 2.6 V for the discharge process can be observed for up to 15 cycles, whereas the charging plateau at 3.6 V shortens progressively. The reduced capacity recovery during charge results in the accumulation of discharge products, and consequently, a steep drop in the discharge (and charge) capacity from the 16$^{th}$ cycle marks the end-of-life the cell. The previously mentioned CCCV protocol was not used for these cells in order to reach the end-of-life stage at an accelerated rate and study their recyclability.

SEM images of electrodes after 21 cycles (Fig 4d) reveal the accumulation of discharge products (EDX analysis in Fig S4). The morphology of the products is clearly different from the first cycle, with a thick layer of solid covering the surface of the electrode, although toroid- and platelet/flower-like particles can also be found. FTIR/ATR reveals (Fig 3e) an increasing $\nu_{O-H}$ peak after 21 cycles.

Two possible mechanisms for the origin of cell death were captured in SEM images: pore-clogging (Fig 4d) and surface passivation (Fig 5a). The electrode surface presents a typical surface pore-clogging, with solid materials (oxygen- rich as depicted by the Fig S5 in supporting information) full-filling the free space in-between nanotubes (Fig 4d), while other areas display nanotubes covered by undissolved solids indicating considerable surface passivation (Fig 5a).

### 2.3. End-of-life electrode treatment

After the cell death, two electrode recovery treatment paths were investigated: (1) reconditioning without further nanotube dispersion and (2) recycling of the CNT material via dispersion and filtration to form new electrodes. For the reconditioning process, an acid and deionized water treatments were investigated, whereas for the CNT recycling process, they had first washed them in abundant deionized water, and then re-dispersed them using the same procedure shown in Fig 1. SEM images show efficient removal of solid deposits by either soaking the aged electrode in a pH = 3 acid solution, or water respectively (Fig 5d-e), and or recycling the CNTs (Fig 5f), (detailed in Experimental Section). EDX data in small magnification over a large area for different electrodes (Fig S5) confirms the disappearance of oxygen-containing species after washing.

Fig 5b shows Raman spectra of cycled electrodes after acid treatment at different pH values. The two characteristic bands (D at 1338 cm$^{-1}$, G at 1550 cm$^{-1}$) correspond to the disorder and in-plane order vibration in the CNT. The ratio between the under-peak area of these bands stays unchanged, indicating no deleterious effect to the bulk structure of the CNT tubes after washing with either hydrochloric or sulfuric acid (pH = 1-4).

XPS data (Fig 5c) similarly shows only negligible increases (<1%) of oxygen and other elements in the washed electrodes compared to the pristine ones, suggesting that the electrochemical reaction and the washing process do not significantly alter the structure or surface of the CNT electrodes.

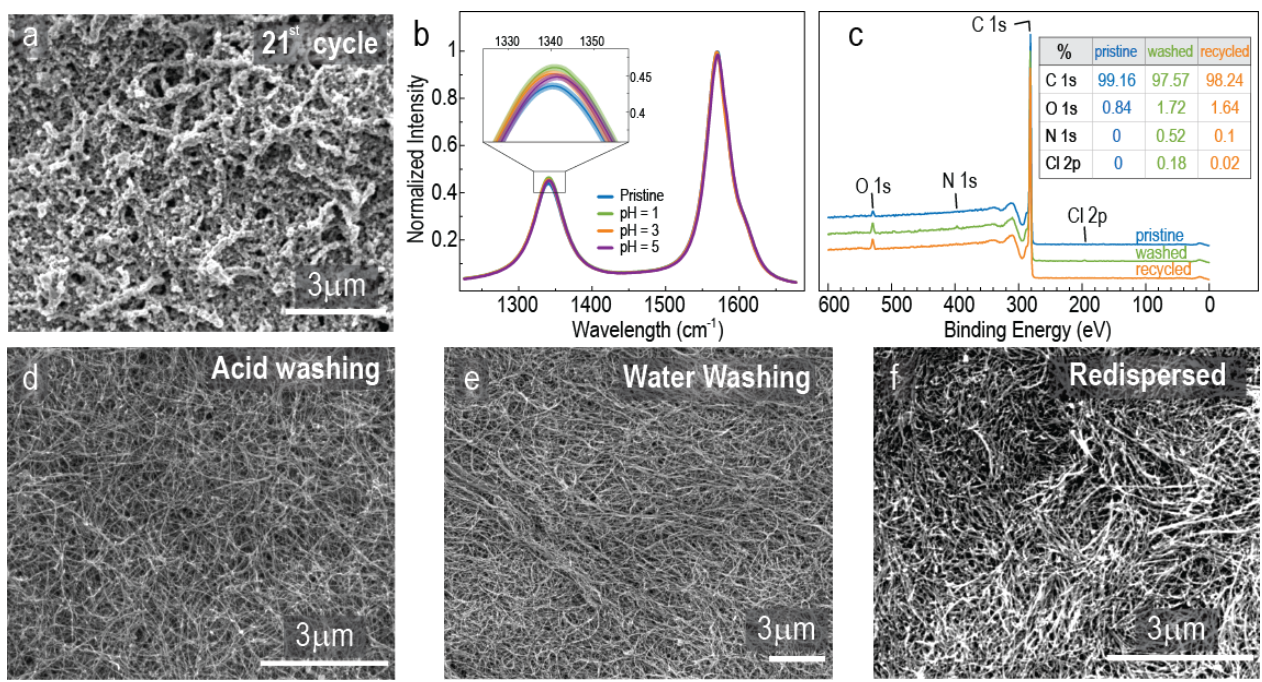

**Figure 5. Characterization of recovered electrodes.** a) SEM image of the surface of an electrode after 21 cycles. b) A Raman study of the acid washing solvent shows no impact of acid onto the bulk CNT. The characteristic CNT D/G bands ratio remains steady at various pH. c) Surface study of the acid washing/recycling CNT by XPS; d-e) SEM images of electrodes post-treatment

### 2.4. Electrochemistry of second-life electrodes

The electrochemical performance of electrodes after reconditioning and recycling were investigated under the same galvanostatic cycling conditions (Fig 6a-f). Both the discharge and charge plateaus remain 2.6V and 3.5 V respectively, with a polarization of ~0.9 V. The efficiency plots showed similar electrochemical performance of the second-life electrodes compared to the initial pristine CNT electrodes (Fig 4a). The capacity retention slowly decreases for the first 15 cycles before an acute drop, due to the fading clogging and passivation mechanisms discussed above.

XRD data of treated electrodes (water-washed and recycled) were analysed prior to and after 1st deep discharge (Fig S6). Reflections in the 30-60 two-theta region indicate the presence of crystalline discharge products ($Li_2O_2$, $LiOH$ and $Li_2CO_3$) in the cycled electrodes. Only minor differences can be observed between the spectra of the pristine, water washed, and recycled electrodes, indicating that the electrochemical products were not impacted by the recondition/recycle processes.

Slightly shorter charging plateaus under 3.6 V were observed when using the acid washed electrodes. Further SEM analysis of the acid washed electrodes after cycling (Fig S7) show significant amounts of flake- and flower-like features alongside toroidal particles, which are typically LiOH crystals. LiOH formation is probably caused by residual protons that are not properly washed-off after acid-treatment, which promotes $Li_2O_2$ hydrolyzation, forming small LiOH crystals.

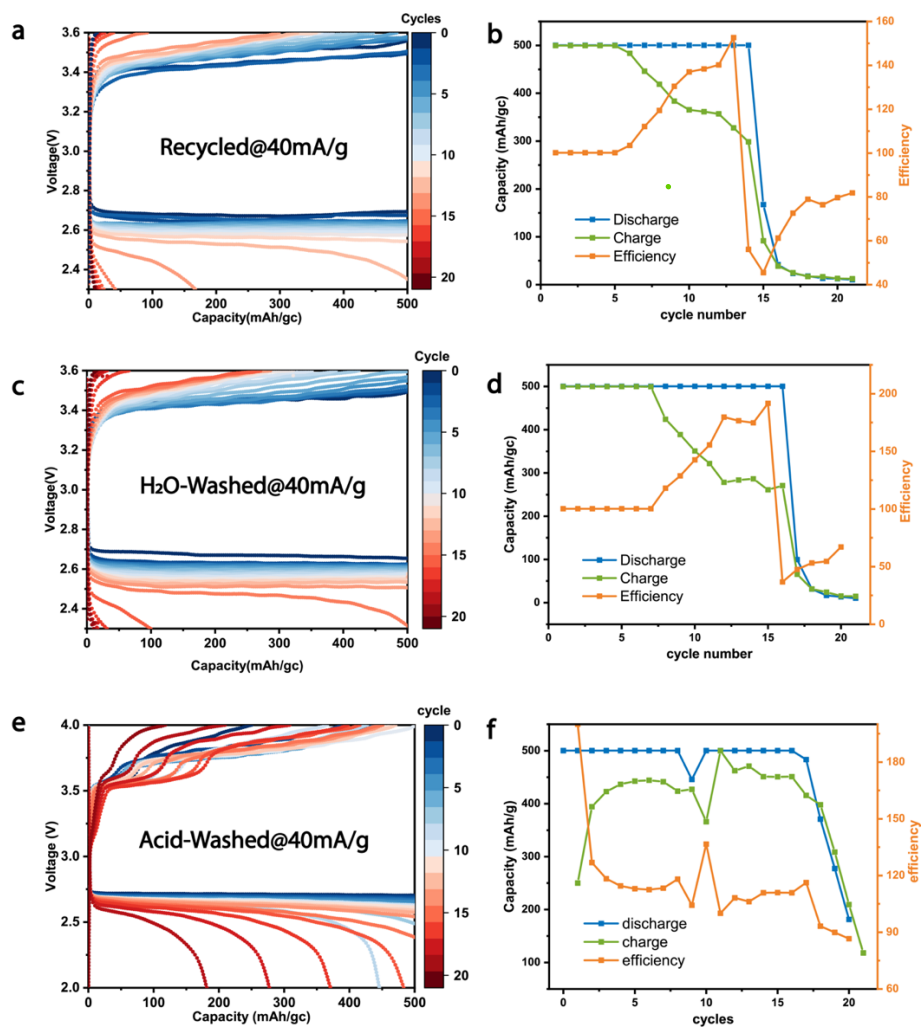

**Figure 6. Electrochemistry of recovered electrodes**. a-b), c-d), and e-f) cycling voltage profiles and columbic efficiency plots of the recycled, water-washed, and pH=3 acid-washed electrodes, respectively.

## 3. Discussion

Our results show that water can be used as efficient washing agent for the removal of solid deposits of spent electrodes in $Li-O_2$ cells, drawing advantages in terms of cost compared with other treatments. This is somehow in contrast to a previous study[45] that suggested that acid washing was essential for discharge product removal. In practice, intense bubbling can be observed when washing the used electrodes by acid (Video SV1). In severe cases, electrodes can be torn into fragments. Whereas using water (Fig S8) makes the bubbling milder, leading to reduced changes of the electrode structure.

The underlying mechanism of the peroxide species removal during washing lies outside the scope of this work, but 2 reactions have been proposed in the LAB literature to account for $Li_2O_2$ chemical decomposition in aqueous media [32,46–48]:

$$Li_2O_2 + 2H_2O \rightarrow 2LiOH + H_2O_2 \tag{1}$$

$$H_2O_2 \rightarrow H_2O + O_2 \uparrow \tag{2}$$

$$2Li_2O_2 + 2H_2O \rightarrow O_2 \uparrow + 4LiOH \tag{3}$$

The discharge products can react in two different ways: the lithium peroxide can be firstly protonated into hydroperoxide and/or into LiOH which further dissolves in the water. Herein, we observe a similar reaction during water and acid-washing. Bubbles could be clearly observed during water-washing of three CNT electrodes (Fig S8), resulting from the disproportionation of hydroperoxide. This process can be accelerated by vacuum pumping above the washing solvent to extract gasses trapped in the pores (MWCNT hydrophobicity drives to wetting/penetrating issue). Regarding the washing efficiency, most lithium salts can be dissolved in water (Table ST1). E.g., the solubility of lithium hydroxide is 127mg/mL, which means small amounts of water are sufficient to wash the amount of product accumulated (<20 mg) in our cells. Whereas acid washing is more efficient at removing the solid deposits (as it increases the solubility of $Li^+$) our results indicate that using water can also efficiently wash discharge products via the two steps reaction-dissolution process. Through the above analysis, if carefully eliminated the protons, which can alter the electrochemical reaction, one can generalize such a recycling process to a broader range of products and battery technologies. For instance, as shown in table ST1, the frequently reported products or by-products in $Li-O_2$ cells (e.g., $Li_2CO_3$, $Li_2O$) are soluble or partially soluble in water. A few of these products are basic (e.g., LiOH) and can be removed with acidic aqueous solutions. More generally, other Metal-$O_2$ batteries such as Na-$O_2$, Zr-$O_2$, and Al-$O_2$ can also benefit from this method as acids can dissolve metal oxides but not carbon matrices.

**Conclusion:**

This work reports a simple method to prepare binder-free self-standing electrodes form cross-linked MWCNT networks, with high porosity and flexibility. Coiled nanotubes electrodes were made from the MWCNT powder by a three steps process: ultrasonication, centrifugation, and vacuum filtration. These free-standing electrodes were tested in LOBs showing capacities of ~1480 mAh/$g_{CNT}$, equivalent to more than 3700Wh/$kg_{CNT}$ of energy density. We demonstrated that these binder-free electrodes can be treated at end-of-life, by efficiently removing adsorbed species using water or acid solutions, by two methods: (a) reconditioning the electrode or (b) reclaiming the carbon nanotube and remaking a new electrode. The reconditioned electrodes were then tested for second-life application, showing almost identical electrochemical performance than when freshly prepared. Furthermore, we suggested routes of undergoing reactions of such recycling method and given an outlook of applicability to other metal-$O_2$ batteries. These findings will be helpful in the development of greener high energy density batteries while reducing recycling costs and environmental impacts.

**Experimental Section:**

### 3.1. Self-standing electrode preparation:

Free-standing binder-free electrodes were prepared using 20 mg of MWCNT (Nanotech Lab.) powder, added to 750 mL of Isopropanol and dispersed using an ultrasonic probe (Sonoplus UW 2200) with intermittent pulses of 200 W for 30 minutes. The MWCNT solution was then centrifuged in 300 mL bottles at 400 rpm for 45 minutes. The clear supernatant solute was vacuum filtrated with the GF/C$^{TM}$ Whatman glassy microfiber filter. The precipitation solution was reused to repeat the procedure a second time. The solution of the second centrifugation became clearer, indicating that the dispersed MWCNT was less concentrated. The filtrated MWCNT was entangled and formed a self-standing porous electrode. Above a MWCNT loading of 2 mg/cm$^2$, the self-standing electrode could be easily peeled off.

The above as-prepared electrodes were dried under vacuum overnight at 120°C. 16 µm thick Celgard (ENTEK 16 µm) and Whatman (Sigma) separators were similarly dried under vacuum at 70°C prior to cell assembly. LiTFSI (Sigma Aldrich), LiNO$_3$ (Alfa Aesar) and dimethylacetamide (Alfa Aesar) were used as received. The water level in the electrolyte was measured at less than 100 ppm by Karl Fischer titration ahead of storing with molecular sieves in an Ar-filled glovebox.

### 3.2. Self-standing electrode reconditioning and recycling

Reconditioning was performed by soaking the cycled electrodes (after cell death) in a hydrochloric acid or water bath (see Video S1) for 30 minutes, inducing bubble formation. The current collector detached from the MWCNT electrodes during the bubbling process. In order to extract the gas trapped in the pores and to impregnate the HCl/H$_2$O into the structure, vacuum pumping was performed above the liquid. For the recycling, self-standing electrodes were washed in HCl/H$_2$O and then in ethanol using a Buchner filter. After weighting, the remaining powder was re-dispersed by a similar process shown in Fig.1 to re-form fresh electrodes (with a smaller 2.5 cm of diameter Buchner funnel).

### 3.3. Electrochemical measurements:

Cell construction was done by directly using the filter of the vacuum filtration process with an added Celgard on the anode side, to slow down oxygen percolation to the lithium foil and avoid occasional short circuit issues from CNT percolation through Whatman filters. 100 µL of electrolyte was used. Electrochemical cycling was performed using a Bio-logic VSP. The Pressure Cell setup was developed following the design described by F. Lepoivre et al.[41]. Details of the e$^-$/O$_2$ ratio calculation can be found in Figure S9. A 2h rest at open-circuit voltage was systematically performed before galvanostatic cycling, allowing thus oxygen diffusion in the electrolyte. A capacity retention comparison with SP@GDL electrodes (Fig S10) was performed using Whatman glassy fiber separators instead of the Celgard separators to limit wetting issues in SP@GDL electrodes.

### 3.4. Post-mortem characterization:

*3.4.1 3D x-ray imaging (nano-CT)*

*Ex-situ* X-ray nano-holotomography[34] acquisitions were performed at the ESRF ID16B beamline [49]. Four tomographic scans, constituted each of 3203 projections, were recorded on a PCO edge 5.5 camera (2560×2160 pixels$^2$) along a 360° rotation with an exposure time of 45 ms per projection using an incident X-ray beam having an energy of 17.5 keV and a high flux of 1.4 $10^{11}$ ph/s. The total acquisition time was approximately 20 minutes per full holo-tomography scan. 3D reconstructions were achieved in two steps: *(i)* phase retrieval calculation using an in-house developed octave script based on a Paganin-like approach using a delta/beta ratio of 303, and *(ii)* filtered backprojection reconstruction using ESRF software PyHST2[50], for a final volume of 64×64×54 µm$^2$ and a voxel size of 25 nm. For tomography measurements, samples have been carved as small tip using carving a Zeiss Palm laser beam and subsequently retrieved thanks to an epoxy-wetted pencil lead. The cycled electrodes were soaked in dimethylacetamide to leach the salt and then dried in a vacuum without heating.

3.4.2 Raman spectroscopy measurements

Dried electrodes were sealed between two glass slides, and 532 nm laser source of 10 mW power with a diaphragm of 50 µm slit was used in the Thermo-Fisher Scientific™ DXR™2 Raman microscope during the acquisition. For each sample in this work, we collected spectra at ten different areas and for each area averaged 32 spectra of 1 second of exposure time.

3.4.3 SEM/EDX acquisition

For the SEM, the cycled cathodes were soaked in the solvent DMA to wash the salt in the dry room. After drying naturally, the DMA in the cathodes, they were transferred from the dry room to the SEM (FEI quanta-200 F) with an airtight sample holder to avoid contamination. Other washed and recycled electrodes were handled under fume hood.

3.4.4 XPS

XPS analysis was carried out using an Escalab 250XI spectrometer from Thermo Fisher Scientific (West Sussex, UK). The instrument was operating in constant analyser energy mode. A monochromatic Al-Kα source (1486.74 eV) and a flood gun for charge neutralization were used, with a spot size of 0.9 mm. Survey scans were acquired using pass energy of 100 eV, using 0.5 eV steps. For narrow scans the number of scans was 10, using pass energy of 20 eV and step size of 0.05 eV. The energetic position of the C 1 s emission line (binding energy of 284.6 eV) was chosen to calibrate the energy scale of all spectra.

**Acknowledgements:**

This research is supported by the French Ministry Higher Education, Research and Innovation. The authors acknowledge the European Synchrotron Radiation Facility for provision of beam time (in-house research time) using the ID16B beamline. They also acknowledge use of the Cambridge XPS System, part of Sir Henry Royce Institute - Cambridge Equipment, EPSRC grant EP/P024947/1and Dr Carmen Fernandez-Posada for XPS data acquisition and processing. IT and and CPG acknowledge funding from the European Research Council (ERC) BATNMR project. A.A.F. acknowledge the European Union's Horizon 2020 research and innovation program for the funding support through the European Research Council (grant agreement 772873, "ARTISTIC" project). A.A.F. acknowledges Institut Universitaire de France for the support.

# Supporting information

**Video SV1.** (Available) Video of violent bubbling during electrode acid washing.

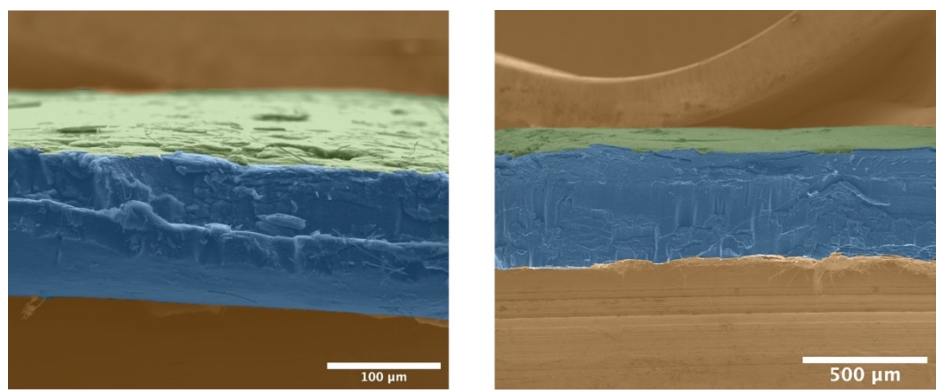

**Figure S1. SEM images showing CNT air electrodes of various thicknesses:** Air-cathodes of different thickness were vertically pinched and imaged in the SEM. The images of self-standing electrodes with thickness of 125 μm (on the left) and 406 μm (on the right) are arbitrarily masked with different colors. The blue, green, and orange colors represents the intersection, surface and background, repectively.

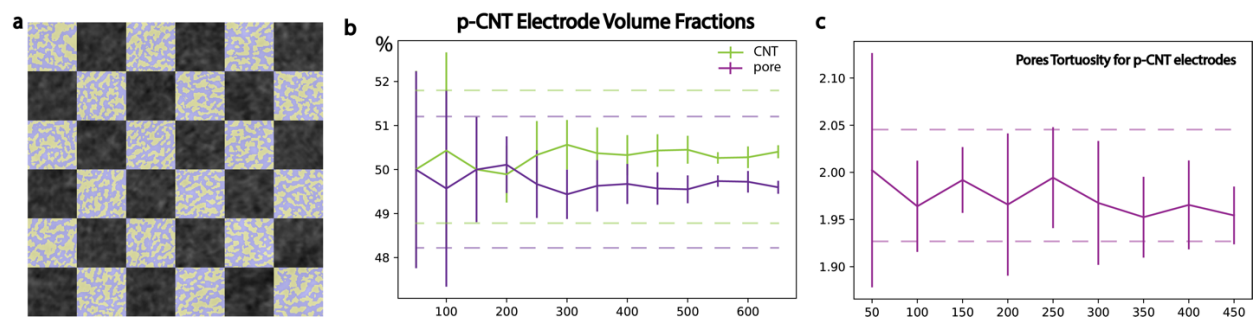

**Figure S2. Segmentation proof and Representative Volume Element (RVE) calculation for both electrodes.** a) a mosaic image with the transparent segmentation mask on the top of tomography slice. b-c) RVE calculation of volume fractions and tortuosity, respectively, the x-axes represent the size of the scanning cube and the y-axes the physical properties (volume fractions for b, tortuosity for c). The dash-lines are plus/minus 3% of uncertainty of the properties for guiding eyes.

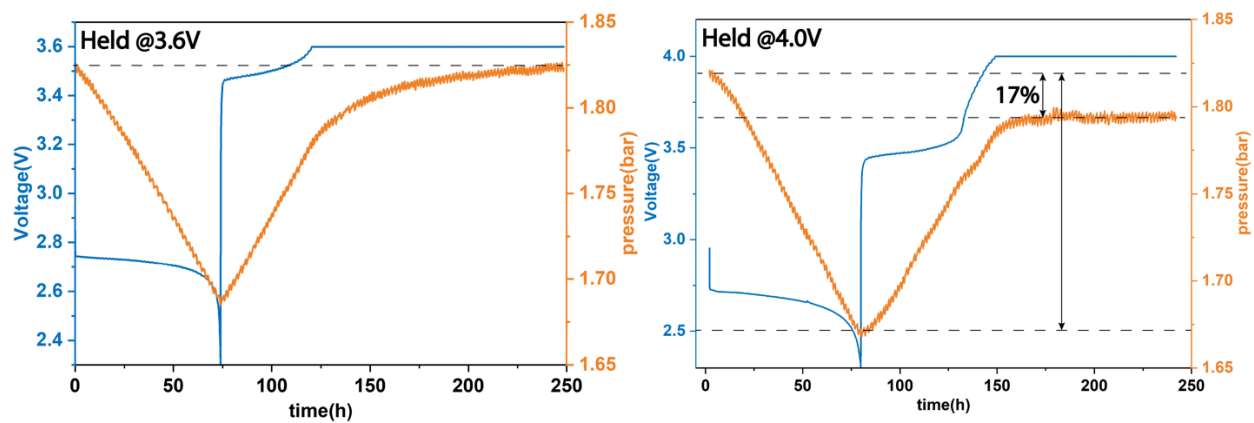

**Figure S3. Comparison of different charging cutoff voltages for the first cycle.** (left) constant current until holding constant voltage at 3.6V and (right) at 4.0V. One can see that holding at 4V does not imply the oxygen evolution reaction, and finally lost 17% of initial pressure.

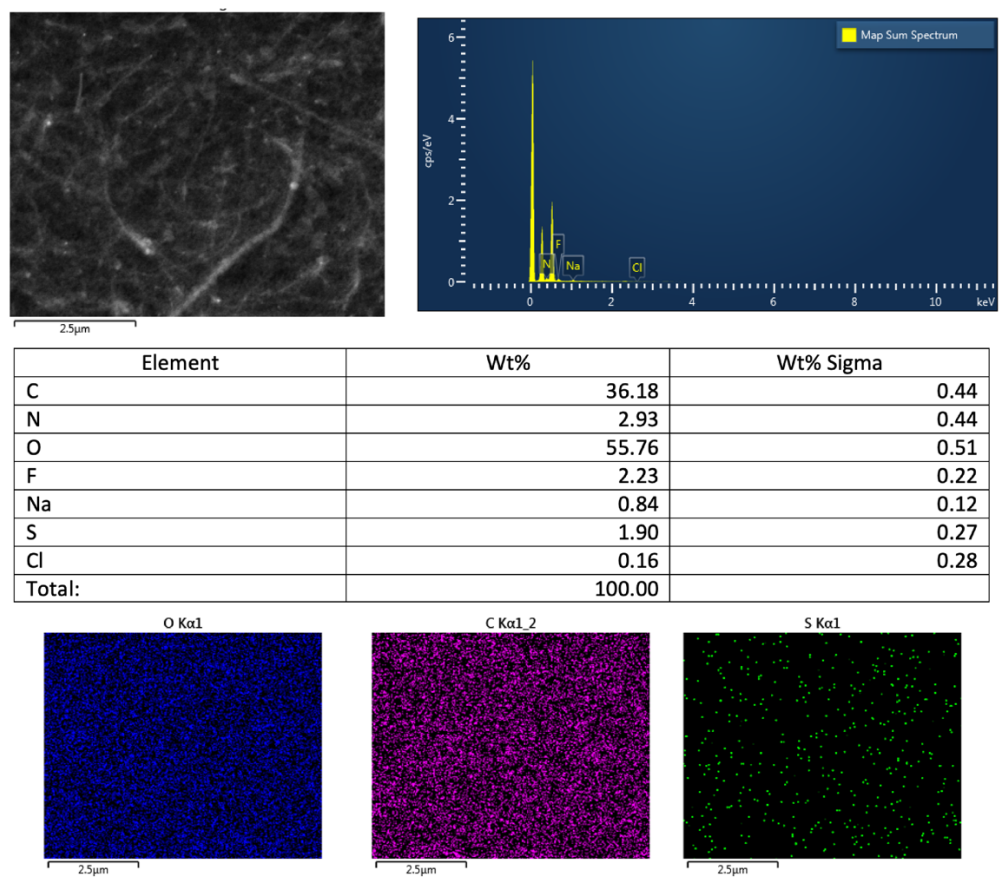

| Element | Wt% | Wt% Sigma |
| --- | --- | --- |
| C | 36.18 | 0.44 |
| N | 2.93 | 0.44 |
| O | 55.76 | 0.51 |
| F | 2.23 | 0.22 |
| Na | 0.84 | 0.12 |
| S | 1.90 | 0.27 |
| Cl | 0.16 | 0.28 |
| Total: | 100.00 | |

**Figure S4.** SEM image and EDS analysis of electrodes after 20 cycles. The upper-left panel is the scanned image of the analyzed zone. The upper-right panel is the EDS spectrum and the middle table is the corresponding element proportions. The three element mappings at the bottom from the left to the right are of the oxygen, carbon, and sulfur, respectively. All three elements are evenly distributed.

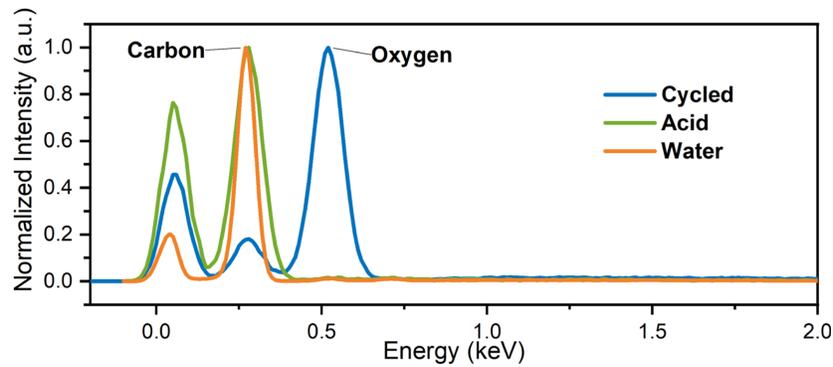

**Figure S5.** EDX/SEM of electrodes after 20 cycles and subsequent acid/water washing. All three spectra are collected in magnification of 2000x enclosing large amount of area to obtain a global comparison.

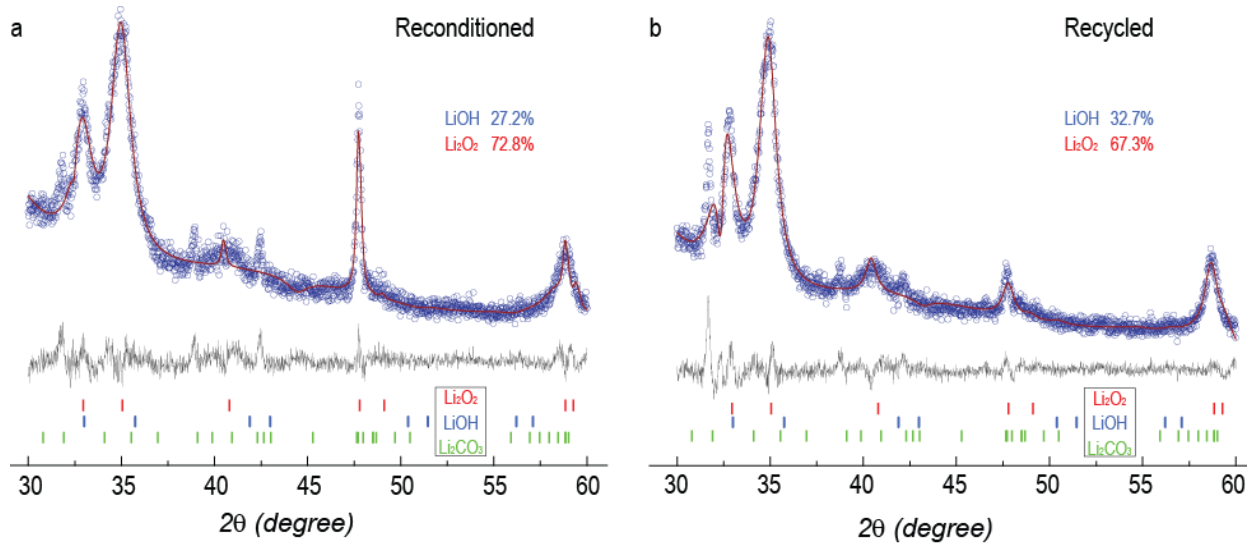

**Figure S6. XRD spectra with Rietveld Refinement comparison** after the deep discharge for a) reconditioned the main peaks positions and are indexed with different colors at the bottom. The residuals are in the middle. b) recycled electrodes, the control pristine one is depicted in Figure 3d. All three electrodes have similar XRD spectra indicating similar cycling products.

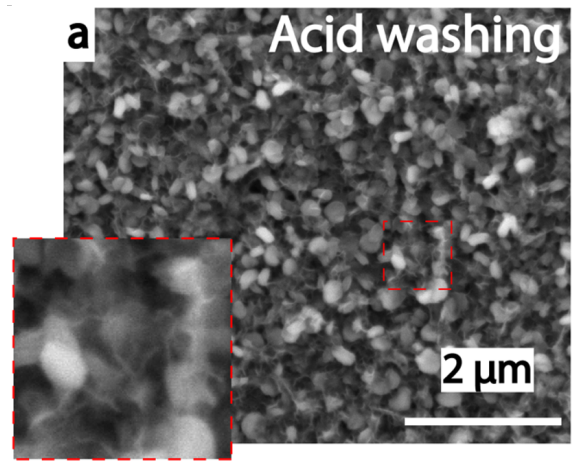

**Figure S7.** SEM image after the 1st discharge of acid-reconditioned electrode. The highlighted zoom shows a close look onto aggregates of lithium peroxide 'dots' and lithium hydroxide flakes with a different size and morphology as the carbon nanotubes.

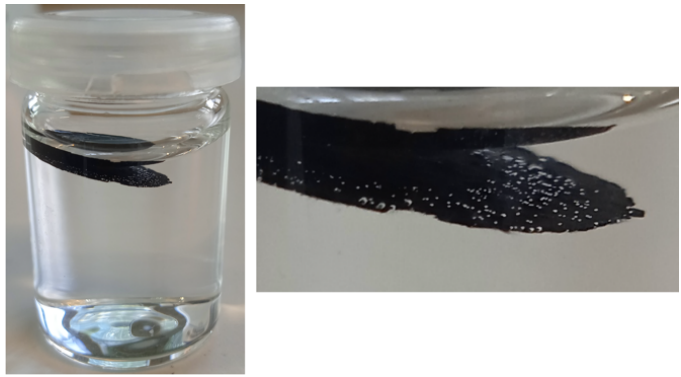

**Figure S8. Gas bubbles are present while using water for washing.** CNT electrodes are heavier than water. Floating electrodes indicates that there is gas in the electrode. To better impregnate the electrode, we made a vacuum above the solution.

Figure S9 shows the Oxygen flushing system and the Pressure-Cell of Collège de France developed by Lepoivre et al.[1]. The pressure captors are calibrated with the GPB 3300 absolute barometer. The Pressure-Cell with the Swagelok was firstly assembled in the Ar-filled glovebox and then replaced the argon with pure oxygen using the flushing system. The flushing is done in three steps: (1) connect the pumping system with the pressure cells with the fast connector component 1 in the left picture and the component 3 in the right picture. (2) pump the left flushing system into vacuum and close the valve to the pump, then slowly open the valve of the pressure cells. The argon gaz is pumped out and rapidly close the pressure cell valve to not let the electrolyte evaporate in low pressure. (3) open the valve of the oxygen to fill the tubing slightly higher than 1.8 bar as indicated in the gauge in the left picture. Re-open the valve of the pressure cell, the pressure of the cell and the tubing are balanced at 1.8 bar. (4) repeat step (2-3) twice, then close the valve of pressure cell and connect it to the VSP for cycling. During the cycling, the blue valve is closed, separating $V_1$ and $V_?$. The corresponding function of the highlighted parts with numbers in Figure S2 are listed in the following:

1. Female quick connect body (airtight without connecting to the male part)
2. Pressure captor OMEGA PX309-030A5V
3. Male quick connect body
4. An insulating ceramic gasket is placed inside the tube for isolating the Swagelok cell from the metallic column to avoid electric perturbation.
5. Metallic tube of bespoke length
6. Swagelok battery

The volume $V_?$ of the column varies in each experiment due to the different compression of the spring. To calculate the electron/$O_2$ ratio, the exact $V_?$ is measured for each cell after cycling by pumping down volume $V_1$ (which is fixed and can be determined beforehand by filling a liquid and measuring the volume of it). The pressure before opening the blue valve in Fig S9 denotes $P_1$.

Then open the valve, and the pressure will drop, and one notes pressure P₂ after it stabilizes. The V? can now be calculated with the expression below.

$$V_? = \frac{V_1 P_2}{P_1 - P_2}$$

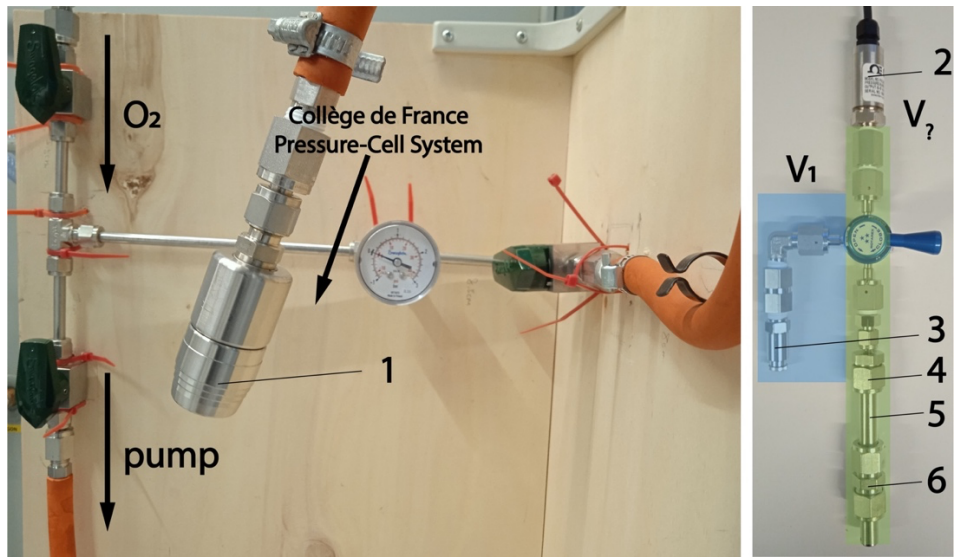

**Figure S9. Digital pictures of the Operando Pressure-Cell System.** (left) the O₂ filling panel (right) the Operando Pressure-Cell System. This closed system was used to full-fill the batteries with oxygen ambience and to measure the pressure during the electrochemical reactions.

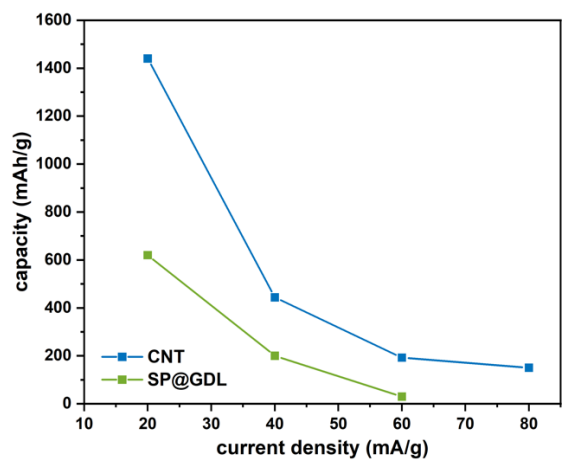

**Figure S10.** Control experiment of capacity in function of the current density with a commercial electrode SP@GDL (Super-P carbon coated onto a Gas Diffusion Layer of large carbon fibers). 50μL extra electrolyte were used for this experiment for better wetting both SP@GDL and pristine CNT electrodes

**Table ST1.** A non-exhaustive list of products and byproducts of LOB documented in the literature.

|  | Reactivity with acid | Reactivity with $H_2O$ | Ref |
|---|---|---|---|
| $Li_2O_2$ | $H_2O_2$, $Li^+$ | $H_2O_2$, LiOH | 46 |
| LiOH | $H_2O$, $Li^+$ | Soluble  127 mg/mL | 32,31 |
| $Li_2CO_3$ | $CO_2$, H2O, $Li^+$ | Mildly soluble  12.9 mg/mL | 8,51 |
| Li-R (strong base) | Protonation | Protonation | 52 |
| LiCl |  | Soluble 842 mg/mL |  |
| $Li_2SO_4$ |  | Soluble 349 mg/mL |  |